\documentclass[%
 reprint,
 amsmath,amssymb,
 aps,
pra,
]{revtex4-2}

\usepackage{graphicx}
\usepackage{dcolumn}
\usepackage{bm}
\usepackage{hyperref}

\begin{document}

\preprint{APS/123-QED}

\title{Complexity of chaotic optical frequency combs}
\thanks{A footnote to the article title}%

\author{Tushar Malica}
\email{tushar.malica@vub.be}
\affiliation{%
 Chaire Photonique, LMOPS, CentraleSup\'elec, 2 Rue Edouard Belin 57070 Metz, France
}
 \affiliation{Universit\'e de Lorraine, LMOPS, 2 Rue Edouard Belin 57070 Metz, France}
\author{Yaya Doumbia}%
 
\affiliation{%
 Chaire Photonique, LMOPS, CentraleSup\'elec, 2 Rue Edouard Belin 57070 Metz, France
}
 \affiliation{Universit\'e de Lorraine, LMOPS, 2 Rue Edouard Belin 57070 Metz, France}
\author{Marc Sciamanna}
\affiliation{%
 Chaire Photonique, LMOPS, CentraleSup\'elec, 2 Rue Edouard Belin 57070 Metz, France
}
 \affiliation{Universit\'e de Lorraine, LMOPS, 2 Rue Edouard Belin 57070 Metz, France}

\date{\today}

\begin{abstract}
The dynamics of a single-mode semiconductor laser induced by the optical injection of a frequency comb are analyzed through the lens of complexity. In particular, the focus is dynamics outside the injection-locking region. Permutation entropy (P.E.) and chaos bandwidth (C.BW.)are used to quantify complexity. Various numerically simulated system outputs across the frequency detuning range and for different injection strengths are considered. Numerically simulated outputs from a single-mode injection system in the absence of comb injection are compared to comment on the origin of the reported complexity. Furthermore, experimental outputs are used to confirm the findings. Despite the presence of periodic dynamics originating from the frequency comb, the presented system generates remarkably high complex outputs with P.E. up to 0.99 maintained over an extended period of observation and long timescales with C.BW. up to 25 GHz.  
\end{abstract}

\maketitle


\section{\label{sec:level1}Introduction}
While injection locking of semiconductor lasers has been studied for over four decades \cite{kovanis1995instabilities,wieczorek2005dynamical,simpson1999phase,simpson1995bandwidth,erneux2009applied, Fortier_Baumann_2019,lin2009nonlinear,desmet2020laser,wu2021optical,quirce2020nonlinear}, nonlinear dynamics exhibited by a single-mode laser diode upon optical injection of a frequency comb is a relatively new topic of interest\cite{desmet2020laser,wu2021optical,quirce2020nonlinear,doumbia2020optical, Fortier_Baumann_2019, doumbia2023wideband}. Various periodic and non-periodic dynamics, including nonlinear wave mixing \cite{shortiss2019harmonic,tistomo2011laser}, selective amplification of the comb \cite{gavrielides2014comb,wu2013selective}, unlocked time-periodic dynamics \cite{shortiss2019harmonic}, and tailoring comb properties by varying injection parameters \cite{duill2017injection,gavrielides2014comb}, have been predicted numerically\cite{doumbia2020optical} and demonstrated experimentally~\cite{doumbia2020nonlinear}. In particular, the system's chaotic nonlinear dynamics have been acknowledged but not investigated deeply.  \\ 
In this letter, we analyze the complexity of numerically simulated system outputs exhibited by a single-mode semiconductor laser induced by the optical injection of a frequency comb. Upon injection, the parameter space region may be classified broadly into injection-locking (IL) and non-injected-locking (NIL) regions\cite{desmet2020laser,wieczorek2005dynamical,doumbia2020optical,doumbia2020nonlinear,quirce2020nonlinear}. The present literature typically has studied this system to optimize the quality of the frequency comb solutions that fall under the IL solutions category\cite{wu2013direct}. IL of a frequency comb results in the selective amplification of the comb lines with a minor shift of the injected laser's frequency position and suppressed unlocked comb lines\cite{doumbia2020nonlinear,doumbia2020optical,wu2013direct}. This has found applications where frequency comb control and manipulation are desired, such as telecommunications\cite{blumenthal2018integrated}, optical metrology\cite{udem2002optical}, and spectroscopy\cite{bernhardt2010cavity}. The NIL region is much less explored. We approach this system from a complexity and entropy standpoint, focusing on chaotic solutions in the NIL region. Gaining an understanding of the nature and trends of nonlinearity of such systems will contribute towards improvement in the performance of applications such as optical chaos communication\cite{cotter2019integrated,bordonalli2015optical}. The system is referred to as a comb injection (CI) system from here onwards. The optical frequency comb (OFC) has a characteristic comb spacing and comb lines. The injection strength and the detuning define the parameter space region. We compare this with the case of single mode injection (SMI) with no OFC injected to comment on qualitative complexity measures using permutation entropy (P.E.) and chaos bandwidth. Finally, we present examples of experimental system outputs to confirm our findings.
\section{\label{sec:level2}Theoretical model}

The model is based on works in~\cite{mogensen1985locking} and used in our previous works as seen in~\cite{doumbia2020nonlinear,doumbia2020optical}. The complex field of the SMI laser, $E_{inj}(t)$, and the injected comb,$E_{ofc}(t)$ having corresponding initial phases as $\phi(t)$ and $\theta(t)$ respectively, are stated as:
\begin{equation}
E_{inj}(t)=E(t)e^{i(2\pi f_0t+\phi(t))}
\end{equation}
\begin{equation}
E_{ofc}(t)=\sum_{j=1}^3 E_j(t)e^{i(2\pi f_jt+\theta(t))}
\end{equation}
where $f_0$ and $f_j$ are frequencies associated with the SMI laser and $j$ number of comb lines injected into the SMI laser. Here, $1\leq j\leq 3$, and so a 3-comb line injected laser is simulated. We limit ourselves to inject 3 comb lines for a realistic study since we are limited experimentally by the resolution photodiode's bandwidth, which in turn, limits the bandwidth measurement of the chaotic signal faithfully, ~\cite{doumbia2023wideband}. Including noise doesn't impact the presented results. The detuning between the injected and the $j_{th}$ comb line is introduced as $\Omega=f_j-f_0$. It should be noted that the detuning in case of comb injection refers with respect to the central comb line. In the case of 3 comb lines, this reference is the second line in the optical spectrum of the injected comb. The initial comb is symmetric, with the same magnitude of amplitude for each comb line ($E'$). The initial phase is set to zero. The zero detuning is calculated from the central injected comb line as well. The differential rate equations for the SMI laser are given as:
\begin{equation}
\frac{dE}{dt}=\frac{G_N}{2}(N(t)-N_{th})E(t)+E'\sum_{j=1}^3 cos(2\pi \Omega t-\phi(t))
\end{equation}

\begin{equation}
\frac{d\phi}{dt}=\frac{\alpha G_N}{2}(N(t)-N_{th})E(t)+\frac{E'}{E(t)}\sum_{j=1}^3 sin(2\pi \Omega t-\phi(t))
\end{equation}

\begin{equation}
\frac{dN}{dt}=R_p-\frac{N(t)}{\tau_s}-G_N(N(t)-N_{th})E(t)^2-\frac{E(t)^2}{\tau_p}
\end{equation}
where $N_{th}=2.91924 \times 10^{24}~m^{-3}$ is the threshold carrier density, $\alpha=3$ is the linewidth enhancement factor, $G_N=7.9\times 10^{-13}~m^3s^{-1}$ is the the differential gain,$\tau_s=2\times 10^{-9}~s$ is the carrier lifetime, and $\tau_p=2\times 10^{-12}~s$ is the photon lifetime. $R_p$ is the pump rate and is 1.2 times the pump rate at the threshold, $R_{th}=1.8 \times10^{33}~s^{-1}$. The parameters are based on previous works on the same topic for which excellent agreement between experiments and numerical results have already been shown~\cite{shortiss2020optical,doumbia2020nonlinear,doumbia2020optical}. The equations are integrated using a fourth-order Runge- Kutta method with a time step equal to 1.2 ps. Numerical simulations are performed for 80 ns. The sampling time is equal to the integration step.

\section{Methodology}
 Permutation entropy (P.E.) is applied to quantify complexity across the range of detuning. This algorithm, developed by Bandt and Pompe~\cite{bandt2002permutation}, is known to measure the predictability of the recurring temporal patterns in a time series through comparison of the relative magnitude of the data points and calculating the probability distribution of user-defined subsets of the time-series. Its extreme robustness to nonlinearity, simplicity and speed of execution make PE an attractive candidate for complexity analysis~\cite{bandt2002permutation,riedl2013practical}. Furthermore, it has been employed in a wide range of applications, including semiconductor laser systems~\cite{soriano2011time,malica2020spatiotemporal,Aragoneses2014}. The algorithm is detailed in~\cite{bandt2002permutation,riedl2013practical} and summarized here. Consider three user-defined parameters, namely, ordinal pattern length ($D$), length of the time series ($N$), and the delay ($\tau$). The ordinal pattern length corresponds to the number of data points extracted from a time series with each data point $\tau$ apart. The delay value is defined as the ratio of the timescale at which one wishes to quantify the complexity to the sampling time ($t_{samp}$). The widely accepted parameter conditions are $3 \leq D \leq7$ and $N >> D$~\cite{riedl2013practical}. Compared to the ordinal pattern length, the significantly higher length of the time series allows the algorithm to construct many subsets of time series, known as the ordinal pattern sets ($\Delta$). A maximum of $D!$ number of ordinal patterns may be constructed. The normalized PE ($\rho_{\tau}$) for a given probability distribution ‘$p$’ associated with ‘$i$’ integral number of ordinal patterns and timescale ‘$\tau$’ is mathematically given by:
\begin{equation}
\rho_{\tau}=\frac{-1}{ln(D!)}\sum^{D!}_{i=1}p(\Delta_i)ln[{p(\Delta_i)]}
\end{equation}
where $0 \leq \rho_{\tau} \leq 1$ with zero signifying complete predictability while one indicates complete stochasticity. To ensure a high-resolution analysis of the numerical simulations, we choose $D$ =5, $N=6 \times 10^4$, $\tau = t_{samp}$ =1.25 ps to track changes in the temporal order~\cite{riedl2013practical}. For experimental outputs, we have a fixed $\tau = t_{samp}$ =10 ps dependent on the oscilloscope, so the optimal values for P.E. analyses were found to be $D$ =6, $N=6 \times 10^4$. It should be noted that the claim of the chosen set of parameters being 'optimal' is based on a trade-off between information extracted to discuss complexity and computation time. It is based on recommendations as stated in \cite {bandt2002permutation,riedl2013practical}. The results are robust with longer time series and ordinal pattern length. \\    
The chaos bandwidth (C.BW. or $\beta$) is defined using the commonly accepted definition of frequency span encompassing 80\% of the RF spectrum energy after subtraction of the noise floor\cite{malica2020spatiotemporal}.\\
It should be noted that the choice of the two figures of merit is made to comment, on both, the qualitative (P.E.) and the quantitative (C.B.W.) nature of chaos. The C.B.W. is an absolute measure, whereas, the P.E. measures the correlation based on the relative amplitude of the data in the ordinal patterns. This is key when the system exhibits similar C.B.W. values and the nuances in the stochastic quality of the dynamical states are highlighted by P.E. rather than C.B.W.~\cite{malica2022high}. 
\section{Results}
\begin{figure}
\centering
\includegraphics [width=\linewidth]{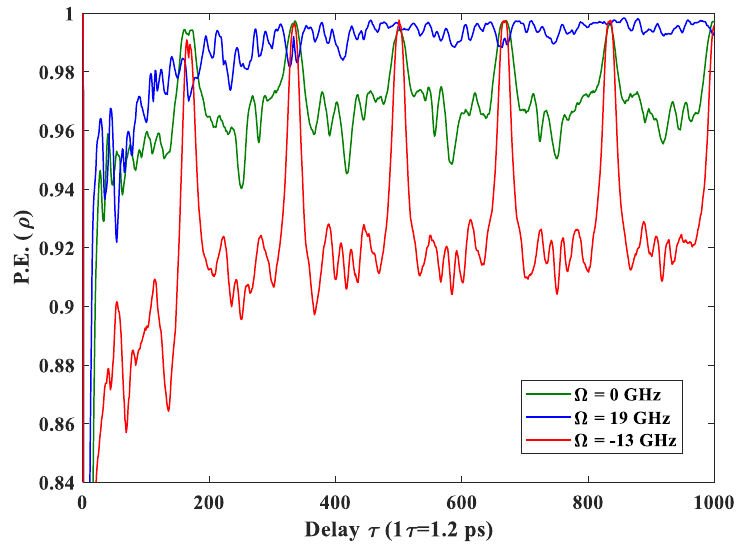}
\caption{P.E. ($\rho$) as a function of delay for simulated output for 3 comb lines, 5 GHz comb spacing, $k_{inj}=0.6$, and $\Omega =$ (a) 0 GHz, $\beta=16.12$ GHz  (b) 19 GHz, $\beta=23.21$ GHz  and (c) -13 GHz, $\beta=19$ GHz.}
\label{fig1}
\end{figure}

Figure \ref{fig1} shows a typical P.E. pattern as a function of delay observed for a system output with varying degrees of chaos present in addition to the comb dynamics at three detuning ($\Omega$) values at strong injection strength ($k_{inj}$). The timescales at which the correlation between system outputs is the lowest correspond to the higher P.E. values and are seen as peaks for $\Omega = $19 GHz and -13 GHz.  The associated timescale of the periodicity at which these peaks occur is the comb spacing (= 5 GHz). Since the periodically pulsed outputs resulting from the presence of the injected OFC are mixed with chaos, the typical dips in P.E. at the periodicity of the comb spacing indicating high predictability are observed to be inverted due to the presence of noise and timing  jitter” \cite{toomey2014complexity}.   It should be noted that the temporal signatures of P.E. here are unlike those reported for the optical-feedback delay systems (OFDS), where P.E. analysis is widely used to investigate timescales at fractional and harmonic values of the round-trip time and relaxation oscillation frequency. The complexity analysis in OFDS primarily focuses on the P.E. values at the round-trip times as an indicator of the quality of complexity. In contrast, the P.E. for OFC injection shows temporal signatures for delays close to the comb spacing, thereby indicating increased predictability at the timescales associated with the comb spacing. Despite that, the OFC injection may increase overall complexity counter-intuitively (P.E.$>$ 0.99) at certain $\Omega$ values, such as the one shown in fig. \ref{fig1} for $\Omega = 19$ GHz where the periodic P.E. patterns associated with the comb lines disappear. Hence, frequency detuning in this system provides us the freedom to navigate within regions of varying complexity.
The C.BW. increases for the three cases shown in fig. \ref{fig1} as the $\Omega$ moves away from zero. The C.BW. at $\Omega$ = 0 GHz is measured at 16.12 GHz. The C.BW. increases more towards the positive detuning as seen in fig.\ref{fig1} for $\Omega$=19 GHz is $\beta$=23.21 GHz. While $\beta$ =19 GHz for a negative $\Omega$ = -13 GHz. In the case of the negative $\Omega$, the presence of temporal signatures is overcome by other processes to still maintain high C.BW. despite the sharp dip in P.E. values at timescales not associated with the comb spacing. The $\Omega$ also increases correlation in the system output, thereby introducing  P.E. patterns that are absent at 0 GHz, especially at higher delay values. \\
\begin{figure} 
\centering
\includegraphics[width=\linewidth]{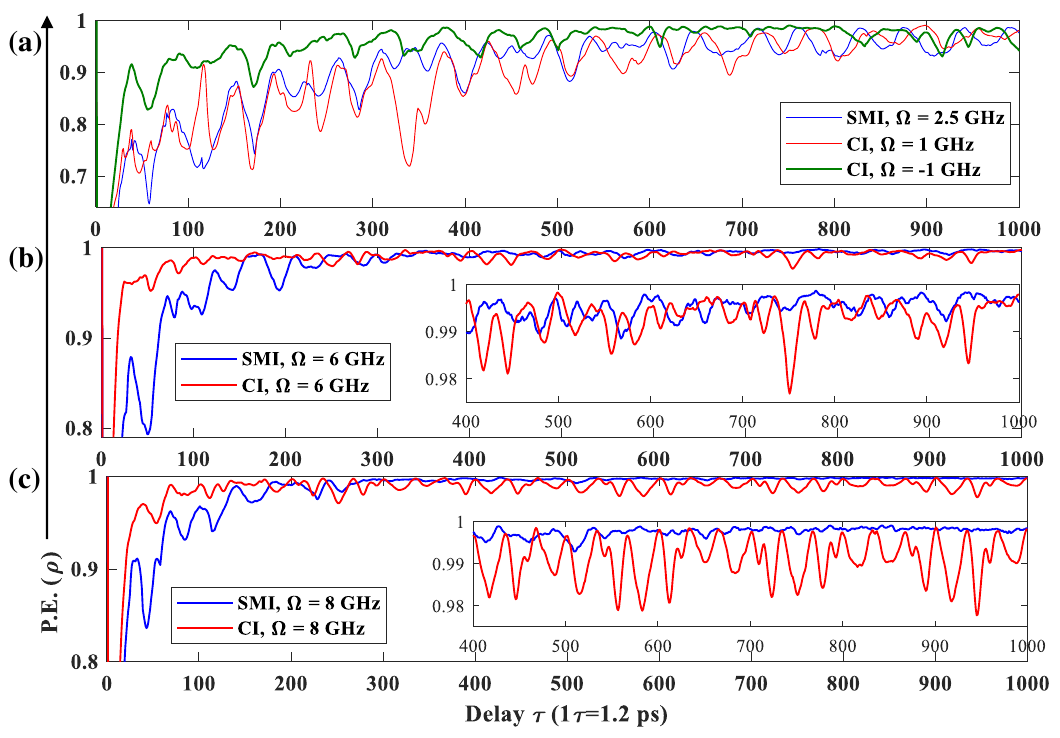}
\caption{Comparison of P.E. as a function of delay ($\tau$) for simulated output for three comb lines, 5 GHz comb spacing, and single mode injection at (a) $k_{inj}=0.1$ (b) and (c) $k_{inj}=0.6$. The C.BW. values are (a) $\beta=14.09$ GHz for $\Omega= 1$ GHz, $\beta=11.19$ GHz for $\Omega= -1$ GHz, and $\beta= 7.48$ GHz for $\Omega=2.5$ GHz; (b) SMI: $\beta= 11.71$ GHz, CI: $\beta= 17.83$ GHz; (c)SMI: $\beta= 14.82$ GHz, CI: $\beta= 18.53$ GHz; Inset showing magnified subset for clarity.}
\label{fig2}
\end{figure}
Figure \ref{fig2} (a) shows P.E. as a function of delay but now for a low $k_{inj}=0.1$. For the same value of $\Omega$ applied to both sides of zero, the complexity is higher at negative $\Omega$ compared to positive $\Omega$ over the extended delay. The P.E. dips observed at timescales associated with the comb spacing, indicating lower complexity, are of lower values for the case of positive $\Omega$. Furthermore, the C.BW. values for both cases are comparable. Compared with SMI, changes in complexity as a function of delay largely remain similar and seem to follow each other for positive $\Omega$ CI and SMI, and deviation from this occurs only at timescales associated with comb spacing. Note that the C.BW. is almost half for SMI compared to the cases of CI at the same $k_{inj}$. Thus, dynamically similar complex outputs from SMI and CI will still differ in their C.BW. measurements, with CI surpassing in performance. 
Similarly, observations for $k_{inj}=0.6$ are shown in fig. \ref{fig2} (b) and (c). The C.BW.  values in both cases are similar for CI. While comparing CI with SMI at the same $\Omega$ and $k_{inj}$, a higher C.BW. always is observed for the CI. Finally, SMI is far more complex at longer timescales, and there is a flattening of the P.E. pattern due to the absence of temporal signatures, as observed in the insets of fig. \ref{fig2} (b)-(c) while the OFC's temporal signature is sustained over longer timescales.

\begin{figure}
\centering
\includegraphics[width=\linewidth]{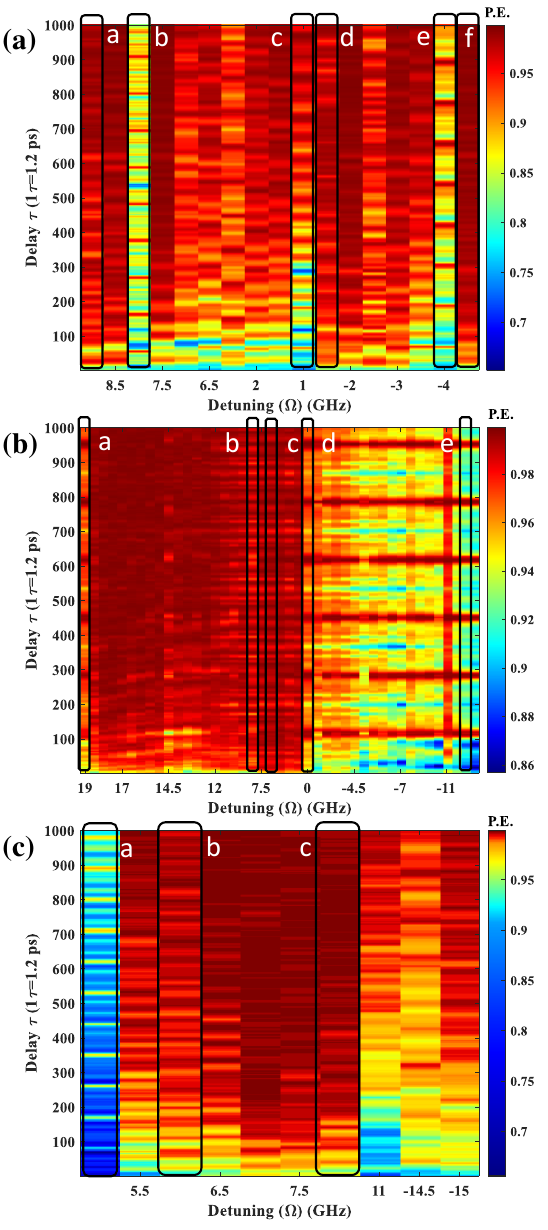}
\caption{Colormap for simulated output showing the delay ($\tau$) as the function of detuning ($\Omega$) in GHz at (a) $k_{inj}=0.1$, and (b) $k_{inj}=0.6$, both for 3 comb lines and 5 GHz comb spacing and (c) single mode injection and $k_{inj}=0.6$. The color intensity indicates the P.E. as seen in the colorbar.}
\label{fig3}
\end{figure}
As we have seen in Figs. \ref{fig1} and \ref{fig2}, the complexity of the chaotic comb can be remarkably high but varies significantly with the injection parameters. Figure \ref{fig3} shows a parameter space colormap for an overall analysis. The vertical axis represents the delay and the horizontal axis represents the $\Omega$ value in GHz. Each color pixel represents a P.E. value with the scale seen as the colorbar. Fig. \ref{fig1} corresponds to cases ``a", ``d", ``e" in fig. \ref{fig3} (b). Fig. \ref{fig2} (a) corresponds to cases ``c" and ``d" in fig. \ref{fig3} (a) for CI and case ``a" in fig. \ref{fig3} (c) for SMI. CI outputs in fig. \ref{fig2} (b) and (c) correspond to cases ``b" and ``c" in fig. \ref{fig3} (b) respectively. SMI outputs in fig. \ref{fig2} (b) and (c) correspond to cases ``b" and ``c" in fig. \ref{fig3} (c) respectively.  
The range of detuning values for CI at lower $k_{inj} = 0.1$, as shown in fig. \ref{fig3} (a) is much smaller than at $k_{inj} = 0.6$. Fig. \ref{fig3} (a) shows special periodic patterns in P.E. and lower value compared to the surrounding region at $\Omega= 1$ GHz, -4 GHz, 8 GHz. These are the values from which the comb would originate and will be discussed later. Figs. \ref{fig1} and \ref{fig2} already show that the C.BW. values at positive $\Omega$ values are much higher for CI. Higher $k_{inj}$ in fig. \ref{fig3} (b) clearly shows the positive $\Omega$ region being more complex with negligible repetitive temporal signatures compared to the negative $\Omega$ region. Figure \ref{fig3} (c) shows a complexity colormap for SMI at $k_{inj} = 0.6$. Weak $k_{inj}=0.1$ in SMI has a restricted range of detuning, which is insufficient to generate a similar colormap. Focussing exclusively on CI maps, an increased $k_{inj}$ also expands the range of detuning values along with operating conditions at detuning values exhibiting high complexity. Additionally, the periodic dips in P.E. occurring with changing delay and corresponding to the temporal signatures associated with the comb characteristics disappear, too. Finally, there is an asymmetry in the system when traversing away from $\Omega$ = 0 GHz with positive detuning, resulting in higher complex system outputs than the negative detuning values where temporal signatures associated with comb characteristics still dominate. It should be noted that comparing CI and SMI shows a larger range of detuning values for CI, and P.E. values are much higher over the entire span of delays. This suggests that CI does, in fact, increase spatiotemporal complexity over the entire range of detuning and thus in the system overall, even though the laser gets injected with a highly ordered structure. Here, the spatial dimension is defined by the delay of the permutation entropy. The benefit of driving the system with a CI for high C.BW., as we already reported in \cite{doumbia2023wideband} combines with an increased chaos complexity compared to SMI.  

\begin{figure} 
\centering
\includegraphics[width=\linewidth]{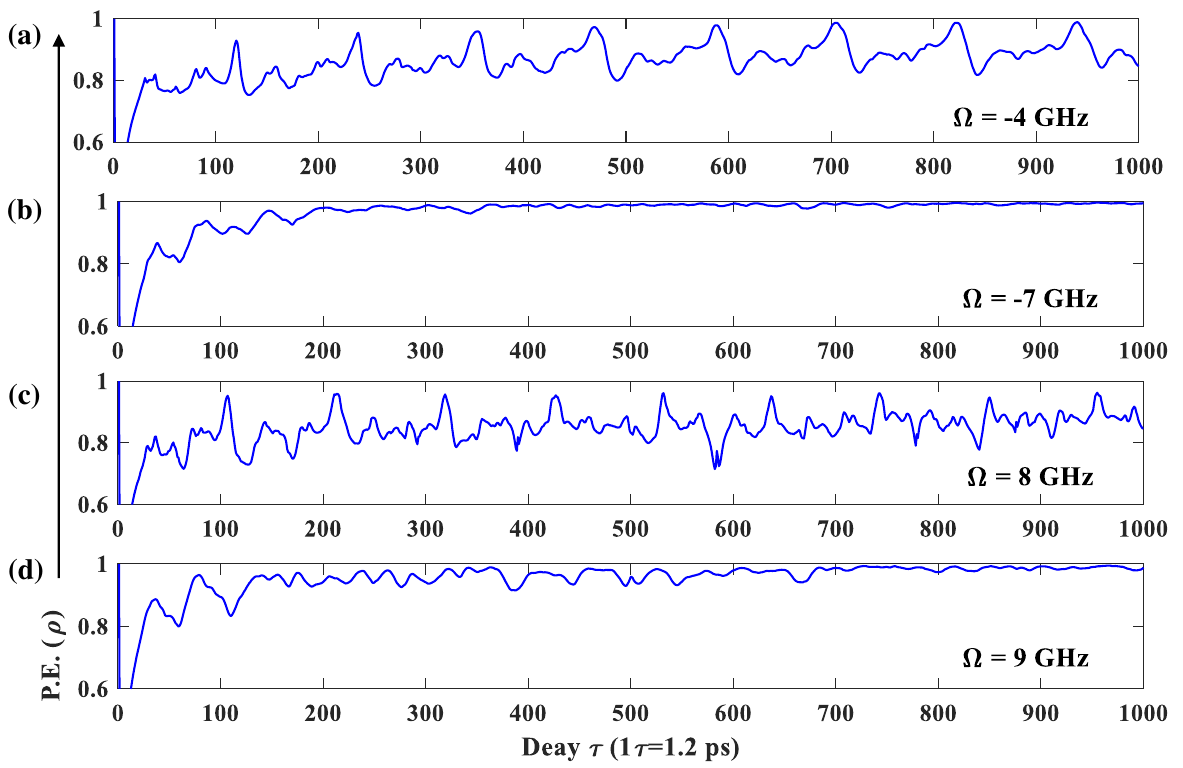}
\caption{P.E. as a function of delay for simulated output for 3 comb lines,5 GHz comb spacing, $k_{inj}=0.1$, and $\Omega =$ (a) -4 GHz,  (b) -7 GHz, (c) 8 GHz, and (d) 9 GHz. The P.E. is calculated for $D=5; N=6\times 10^4$.}
\label{fig4} 
\end{figure}

\par Figure \ref{fig4} shows specific cases of P.E. as a function of delay taken from fig. \ref{fig3} (a) . Figs. \ref{fig4} (a) and (c) correspond to cases ``e" and ``b" indicated in fig. \ref{fig3} (a) respectively. These points in the parameter space show where the comb dynamics emerge. The presence of periodic patterns associated with the frequency comb is reflected in the repetitive P.E. patterns here. The correlation between the data points of the digitized system outputs is maintained as the timescale of observation gets longer. Figs. \ref{fig4} (b) and (d) correspond to cases ``f" and ``a" indicated in fig. \ref{fig3} (a) respectively. These are typical P.E. plots with P.E.$>$ 0.99, where the correlation between the data points of the digitized system outputs is negligible, even at longer timescales. One can conclude that sweeping detuning value at fixed $k_{inj}$ increases complexity in system outputs and, hence, better P.E. values. These results are enhanced with P.E. closer to what one might expect in the case of Gaussian white noise as the delay is increased with the complete flattening of the P.E. plots, as seen in fig. \ref{fig4} (b) and (d) at $\tau > 300 $. For the comb-injected case, our past work in \cite{doumbia2020nonlinear,doumbia2023wideband} and current findings suggest that the statistically measured chaos bandwidth deteriorates from negative to positive values of detuning. For positive detuning, the area of the chaotic signal is limited to low injection strength and the chaos disappears when increasing the injection strength. The complexity, i.e., the measured unpredictability (or stochasticity) as seen in fig. \ref{fig3} for the case of comb injection is overall higher at positive detuning values than the negative detuning values. Furthermore, the nonlinearity is higher at positive values of detunings with low correlation of system outputs observed at shorter as well as longer timescales. On the other hand, negative detunings do show periodicity at timescales corresponding to the comb properties. These timescales are not prominent for positive values of detuning due to the aforementioned nonlinear behaviour.  

\begin{figure} 
\centering
\includegraphics[width=\linewidth]{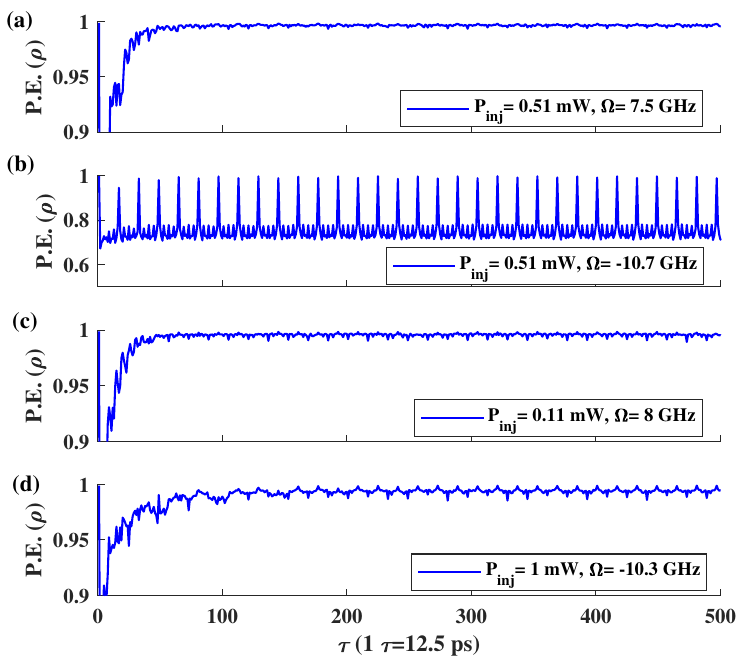}
\caption{Experimental system outputs captured at 12.5 ps sampling time using an oscilloscope of 36 GHz bandwidth showing effects of chaos using a frequency comb with 5 GHz comb spacing and 3 comb lines. (a)-(d): P.E. as a function of delay. Operating conditions are (a):$\Omega$ = 7.5 GHz, $P_{inj}=0.51 mW$, $\beta$ = 18.5 GHz; (b) $\Omega = $-10.7 GHz, $P_{inj}=0.51 mW$, $\beta$ = 22.3 GHz; (c)$\Omega$ = 8 GHz, $P_{inj}=0.11 mW$t, $\beta$ = 17.07 GHz; and (d)$\Omega$ = -10.3 GHz, $P_{inj}= 1 mW$, $\beta$ = 21.6 GHz. The P.E. is calculated at conditions comparable to simulations, i.e., $m=6;t_{samp}=12.5ps; N=48\times 10^3$ .}
 \label{fig5}
\end{figure}
\par Figure \ref{fig5} shows a variety of experimental outputs captured using the experimental setup previously reported by us in \cite{doumbia2020nonlinear}. Note that the injection power ($P_{inj}$) cannot be accurately converted to $k_{inj}$ used in numerical simulations. Also, the P.E. pattern remains unchanged at longer delay values. Therefore, we show a variety of cases. An ideal case with P.E. as a function of delay has an almost flat top, as seen in fig. \ref{fig5} (a). This is similar to the case of white Gaussian noise, suggesting almost complete stochasticity. The temporal signatures from the comb may gain prominence upon sweeping parameters, as seen in fig. \ref{fig5} (b). No pattern is visible in the P.E. with the resulting chaos of optimal complexity despite the chaotic dynamics achieved by the injection of the comb and associated repetitive dynamics. However, this is still not devoid of overall high complexity with the range of P.E. values across the delay range being upwards of 0.99 despite the presence of periodic patterns.

\par The comparison between fig. \ref{fig5} (a) and (b) show that, similarly to the theoretical predictions, positive detunings lead to slightly higher complexity than the observations found at negative detunings. At an even lower injection power is seen in fig. \ref{fig5} (c), one can still see some periodic patterns associated with the comb spacing even though overall complexity remains high, similar to fig. \ref{fig5} (a). This observation remains consistent at high injection power as seen in fig. \ref{fig5} (d). Therefore, we see results from simulation and experimental findings commensurate regardless of whether injection power is high or low and detuning value is positive or negative. Lastly, the presence of a comb introduces periodicity in the system. It might lower complexity at certain timescales, but the overall complexity remains high and comparable to values achieved using OFDS \cite{malica2020spatiotemporal, malica2022high}. 
 
\section{Conclusions} In conclusion, a semiconductor laser induced by the optical injection of a frequency comb exhibits remarkably high complex system output regardless of periodic comb dynamics. The complexity is not limited to the measurement of the chaos bandwidth but also the correlation between digitized system output. This aforementioned correlation is low, sustained over time, and observed at longer and shorter timescales. Depending on the experimental variables such as the frequency detuning of the frequency comb lines with respect to its central frequency and the injection strength, one may traverse across different regions of complexity, including regions exhibiting high permutation entropy and chaos bandwidth. Additionally, there are cases where only one of these figures of merit may be high. Some examples shown prove the presence of operating conditions where the complexity of comb injection is comparable to the case of single mode injection but with a significantly higher chaos bandwidth owing to the high chaotic components generated by the comb injection. Therefore, injection of the comb is observed to improve the laser chaos performances when compared to the case of single mode injection. Hence, optical injection of the combs is an extremely promising approach to generate optical chaos with optimal bandwidth and complexity properties and suggests superior performance in chaos-based applications.

\begin{acknowledgments}
We wish to acknowledge the support of Chaire photonique, Ministere de l’Enseignement Superieur de la Recherche et de l’Innovation, Region Grand-Est,  Departement Moselle, European Regional Development Fund (ERDF), Airbus GDI Simulation, CentraleSupelec, Fondation CentraleSupelec, Fondation Supelec,and Eurometropole metz.
\end{acknowledgments}


\bibliography{References}

\end{document}